\documentclass[10pt,conference,letterpaper]{IEEEtran}
\IEEEoverridecommandlockouts
\usepackage{graphicx}
\usepackage[cmex10]{amsmath}
\usepackage{epstopdf}
\usepackage[short]{optidef}
\usepackage[utf8]{inputenc}
\usepackage[english]{babel}
\usepackage{color}
\usepackage{epstopdf}
\usepackage{commath}
\usepackage{algorithm}
\usepackage[caption=false]{subfig}
\usepackage{cite}
\usepackage{amssymb}
\usepackage{mathtools}  
\usepackage{eucal}
\usepackage{enumitem}
\newcommand\blfootnote[1]{%
		\begingroup
		\renewcommand\thefootnote{}\footnote{#1}%
		\addtocounter{footnote}{-1}%
		\endgroup
}

\title{Trajectory Optimization in UAV-Assisted Cellular Networks under Mission Duration Constraint}

\author{
    \IEEEauthorblockN{Md Moin Uddin Chowdhury\IEEEauthorrefmark{1}, Eyuphan Bulut\IEEEauthorrefmark{2}, Ismail Guvenc\IEEEauthorrefmark{1}}
    \IEEEauthorblockA{\IEEEauthorrefmark{1}Department of Electrical and Computer Engineering, North Carolina State University, Raleigh, NC, USA\\
    \IEEEauthorblockA{\IEEEauthorrefmark{2}Department of Computer Science, Virginia Commonwealth University, Richmond, VA, USA}
        \{mchowdh, iguvenc\}@ncsu.edu, ebulut@vcu.edu}
}

\vspace{-1cm}
\begin{document}
\vspace{-.05cm}	
\pdfoutput=1
\maketitle
\begin{abstract}
In this paper, we address the problem of finding the optimal trajectory for an unmanned aerial vehicle (UAV) for improving the wireless coverage of a terrestrial cellular network. In particular, we consider a UAV that is tasked to travel from one point to another within a given time constraint, and it can also simultaneously assist the cellular network by providing wireless coverage during its mission. Considering an interference limited downlink of a cellular network, we formulate an optimization problem for maximizing the proportional-fair (PF) data rate of the cellular network and explore dynamic programming (DP) technique for finding the optimum UAV trajectory. We also explore the optimal UAV trajectories associated with maximum sum-rate and 5th percentile spectral efficiency (5pSE) rate and compare the capacity and coverage performance of the three approaches. Numerical simulations show that the maximum sum-rate trajectory provides the best per user capacity whereas, the optimal PF trajectory yields higher coverage probability than the other two trajectories. The optimal trajectories are generally infeasible to follow exactly as the UAVs can not take sharp turns due to kinematic constraints. Hence, we generate smooth trajectories using Bezier curve.
\end{abstract}

\section{Introduction}

\blfootnote{This research was supported by NSF under the grant CNS-1453678.} Use of unmanned aerial vehicles (UAVs), also known as drones, is recently finding various civilian applications 
such as aerial surveying, delivery of merchandise and medical supplies, and search/rescue operations~\cite{Valavanis,rui1}. Thanks to their battery power management techniques, the ability to harvest solar energy, and the capability to increase network capacity, UAVs can also be deployed as aerial base stations. In this context, UAVs have a potential in revolutionizing the future of broadband communication networks. 

While dedicated UAV base stations (BSs) can be deployed by service providers for assisting cellular networks, it may also be possible to take advantage of other UAVs that are tasked to travel from one place to another. Such UAVs may for example include (but are not limited to) delivery drones, such as those that may be used by Amazon, which should deliver merchandise to a target location within a specific time window. In this paper, we consider a scenario where a UAV should travel between two points under a specific time constraint, and can assist in providing wireless connectivity of an underlying cellular network in the meantime. A  key challenge here then becomes the optimal design of the UAV's trajectory, in order to provide the best wireless service to users. 

UAV trajectory design considering an underlying cellular network has recently been studied in the literature. For instance, \cite{Bulut} uses dynamic programming (DP)~\cite{dp} to find the optimal UAV trajectory while maintaining good connectivity with cellular  BSs. In~\cite{rajeev}, authors consider DP technique to optimize the weighted sum-rate of users in a wireless network, while~\cite{gesbert} considers landing spots to trade-off throughput with battery power. In both works, authors did not consider the presence of other cellular base stations and did not explore fairness among users. The main goal in~\cite{rui1} is to maximize the minimum throughput of users in a multi-UAV enabled network, while~\cite{rui2} explores the problem of minimizing the mission completion time while ensuring good link quality to enable multicasting via trajectory optimization. Energy efficient trajectory optimization using sequential convexification techniques is discussed in~\cite{rui3}. In \cite{saad}, deep reinforcement learning is used to generate trajectory with an aim to reduce interference, while \cite{jiang} proposes a dynamic UAV heading adjustment algorithm to optimize the ergodic sum-rate of an uplink wireless network. UAV enabled communication system is also studied in \cite{merwaday}, where optimum UAV locations as well as  interference management parameters are solved. 

In this paper, we formulate the UAV trajectory optimization problem between two points to maximize the proportional fair (PF) rate of an in-band downlink cellular network where, the locations of the ground users (UE) and macro base
stations (MBS) follow homogeneous Poisson point processes (PPPs). As the problem is difficult to solve in general, we use dynamic programming (DP) to obtain the optimal path. We also exploit the same technique for optimizing Max sum-rate and fifth percentile spectral efficiency rate trajectories. After getting the trajectories of above three methods, we study and compare the capacity and outage probability performances associated with each of them. From the numerical results, the following insights are obtained: (i)  the Max sum-rate trajectory is found to be more per user capacity efficient than the other two approaches, and (ii) PF rate trajectory provides the best coverage probability. Moreover, we use Bezier curve to smooth the generated trajectories and analyze the capacity and outage probability performances of smooth trajectories.
\section{System Model}\label{Sec2}

Consider a UAV that is flying at a fixed height $H$ with maximum speed of $V_{\mathrm{max}}$ in a suburban environment. The mission of the UAV is to fly from a start location, $L_{\rm s}$ to a final destination point $L_{\rm f}$ within a fixed time $T$ on an area of $\mathcal{A}$ square meters. Let us consider $(x(s),y(s),H)$ and $(x(f),y(f),H)$ to be the 3D Cartesian coordinates of $L_{\rm s}$ and $L_{\rm f}$, respectively. The time-varying horizontal coordinate of the UAV at time instant $t$ is denoted by $r(t)=[x(t),y(t)]^\intercal \in \mathbb{R}^{2\times1}$ with $0\leq t\leq T$. The minimum time required for the UAV to reach $L_{\rm f}$ from $L_{\rm s}$ with the maximum speed ${V_{\mathrm{max}}}$ is given by ${T_{\mathrm{min}}} = \frac {\sqrt{(x(s)-x(f))^2+(y(s)-y(f))^2})}{V_{\mathrm{max}}}$. The UAV's mobility is modeled as, $\dot{x}(t)=v(t)\cos\phi(t)$ and $\dot{y}(t)=v(t)\sin\phi(t)$,
where $\dot{x}(t)$ and $\dot{y}(t)$ are the time derivative of $x(t)$ and $y(t)$, respectively, $v(t)$ is the velocity and $\phi(t)$ is the heading angle (in azimuth) of the UAV.
Let us also assume that there are $M$ MBSs and $K$ static UEs in the area. 
The set of the UEs is denoted as $\mathcal{K}$ with horizontal coordinates  $\textbf{w}_k=[x_k,y_k]^T \in \mathbb{R}^{2\times1}, k\in \mathcal{K}$. The MBS and UE locations follow two identical and independent PPPs. The MBSs transmit with omni-directional antennas and each UE connects to the strongest MBS or the UAV.

We consider sub-6 GHz band and Okumura-Hata path loss model (OHPLM) for all communication links, as it is more relevant for a terrestrial environment where base-station height does not vary significantly~\cite{hata}. We also assume that the network is interference limited, where thermal noise power at a receiver is presumed to be negligible compared to the interference power. 
The Doppler spread stemming from the UAVs mobility is considered to be compensated at the receivers. The path loss (in dB) observed at UE $k \in \mathcal{K}$ from MBS $m$ and the UAV at time $t$ is given by, $\xi_{\mathrm{k,m}}(t)=A+B\log_{10}({d_{\mathrm{k,m,t}}})+C$ and $ \xi_{\mathrm{k,u}}(t)=A+B\log_{10}({d_{\mathrm{k,u,t}}})+C$. Here, ${d_{\mathrm{k,m,t}}}$  and ${d_{\mathrm{k,u,t}}}$ are the Euclidean distances from MBS $m$ to user $k$ and from UAV to user $k$ at time $t$. $A$, $B$, and $C$ are the factors dependent of the carrier frequency $f_c$ and antenna heights \cite{hata}. Then, the received power at user $k$ from  MBS $m$  at time $t$, can be calculated as ${S_{\mathrm{m,t}}}=\frac{{P_{\mathrm{mbs}}}}{10^{\xi_{\mathrm{k,m}}(t)}/10}$. Similarly, the received power at user $k$ from  the UAV at time $t$, can be calculated as ${S_{\mathrm{u,t}}}=\frac{{P_{\mathrm{uav}}}}{10^{\xi_{\mathrm{k,u}}(t)}/10}$. During each $t$, a UE connects to either its nearest MBS or the UAV, whichever provides the best signal-to-interference ratio (SIR). Assuming round-robin scheduling, we can express the achievable data rate per unit bandwidth (bps/Hz) of user $k$ at time $t$ using Shannon's capacity  as: 
\begin{align}
{R_{k}(t)}=\frac{\log_{2}(1+{\gamma_{k}(t))}}{{N_{\mathrm{ue}}}},
\end{align}
where ${\gamma_{k}(t))}$  is the instantaneous SIR of $k$-th user and ${N_{\mathrm{ue}}}$ is the number of users in a cell.


\section{UAV Trajectory Optimization}

Considering the system model in Section~\ref{Sec2}, the logarithmic sum rate of the network at time  $t$ can be expressed as:
\begin{equation}
    C(t)= \sum_{k=1}^K \log_{10} {R_{k}(t)},
    \label{eq}
\end{equation}
which is known to correspond to the proportional fair rate of the network. Now we can formulate our trajectory optimization problem over the total mission duration of the UAV as:  
\begin{maxi!}	
	  {x(t),y(t)}{\int_{t=0}^{T}C(t)}{}{}\label{Eq3}
	  \addConstraint{\sqrt{\dot{x}(t)^2+\dot{y}(t)^2}}{\leq {V_{\mathrm{max}}}, } \; { t \in [0,T]}
	  \addConstraint{(x(0),y(0)) =}{(x(s),y(s))}{}
	  \addConstraint{(x(T),y(T)) =}{(x(f),y(f))}{}
\end{maxi!}
Here, (3b) ensures that velocity of UAV does not exceed the maximum limit, while (3c) and (3d) set the initial and final location of the mission. For finding the optimal trajectory that corresponds to the maximum sum-rate, we can just exclude the $\log_{10}$ term in \eqref{eq}. 
The maximization problem provided above is a non-convex problem which is difficult to be solved optimally in general.

\section{Dynamic Programming Technique}
In this section, the optimization problem in~\eqref{Eq3} is discretized to obtain approximation of the optimal trajectories. The time period $[0,T]$ is divided into $N$ equal intervals of duration $\delta=T/N$ and is indexed by $i=0,....,N-1$. The value of $N$ is chosen so that UAV's position, velocity, and heading angle can be considered constant in an interval. The rate of UE $k$, ${R_{k}(i)}$ will be dependent on the distance between the UE and the horizontal position of the UAV at time interval $i$. Then, the discrete-time dynamic system can be written as:
\begin{equation}
    \mathbf{s}_{i+1}=\mathbf{s}_{i}+ f(i,\mathbf{s}_{i},\mathbf{u}_{i}),\quad  i=0,1,...,N-1,
\end{equation}
where $\mathbf{s}_{i}=[x_i\;y_i]^T$ is the state or the position of the UAV and $\mathbf{u}_{i}=[v_i\;\phi_i]^T$ stands for the control action i.e., velocity and the heading angle, respectively, in the $i$th time interval.  By taking control action at each interval $i$, the UAV will move to next state and it will achieve cost for taking that specific control action. Starting with initial state $s_{0}=[0,0]^T$, the subsequent states can be computed using,
\begin{equation}
    f(i,\boldsymbol{s}_{i},\boldsymbol{u}_{i})= \begin{pmatrix}
  v_i\cos{\phi_i} \\
  v_i\sin{\phi_i}\\
\end{pmatrix}.
\end{equation}
The optimal cost can be calculated recursively using Bellman's equations by moving backwards in time as follows~\cite{dp}, \cite{rajeev}:
\begin{equation}
    J(\mathbf{s}_{i})=\max_{\mathbf{u}_{i}}\sum_{k=1}^K \log_{10} {R_{k}(i)}+ J(\mathbf{s}_{i+1}),\quad  i=N-1,..,0,
\end{equation}
where $J(\mathbf{s}_{i})$ is the cost associated with state $i$, and the terminal cost $J(\mathbf{s}_{N})$ is the cost when the UAV reaches the position $[x(f)\; y(f)]$.

\section{Numerical Results}

In this section, we numerically obtain the optimal trajectories of the UAV by applying DP~\cite{dp}. We consider a square region of dimensions 1~km by 1~km where there exists a cellular network operating at $1.5$~GHz. The number of randomly distributed MBSs ($N_{\rm MBS}$) can be $4,~5,~6$, while we consider $100$ randomly distributed UEs in the area. The transmit power of MBSs and UAV are assumed  to be 46 dBm and 30 dBm, respectively. The maximum velocity ${V_{\mathrm{max}}}$ is considered to be $17.7$~m/s with $\delta=8$~seconds. The height of UAV and BS is $120$~m and $30$~m, respectively, while the UE height is $2$~m. 
We divide both $x$ and $y$ coordinates with a step of $100$~m, which gives us $121$ unique geometric positions where we assume UAVs can modify their trajectory. As we have segmented all possible states into finite discrete geometrical positions, we consider the following control actions on the map:
\begin{equation}
    \mathbf{u} \in \Bigg\{\begin{bmatrix} 0 \frac{m}{s}\\ \\0 \end{bmatrix},\begin{bmatrix} 12.5 \frac{m}{s}\\ \\\theta \end{bmatrix},\begin{bmatrix} 17.7 \frac{m}{s}\\ \\\theta+\frac{\pi}{4} \end{bmatrix} \Bigg\},
\end{equation}
where, $\theta \in \{0,\frac{\pi}{2},\pi,\frac{3\pi}{2} \}$.

\begin{figure}[t]
		\centering
		\subfloat[]{
			\includegraphics[width=0.8\linewidth]{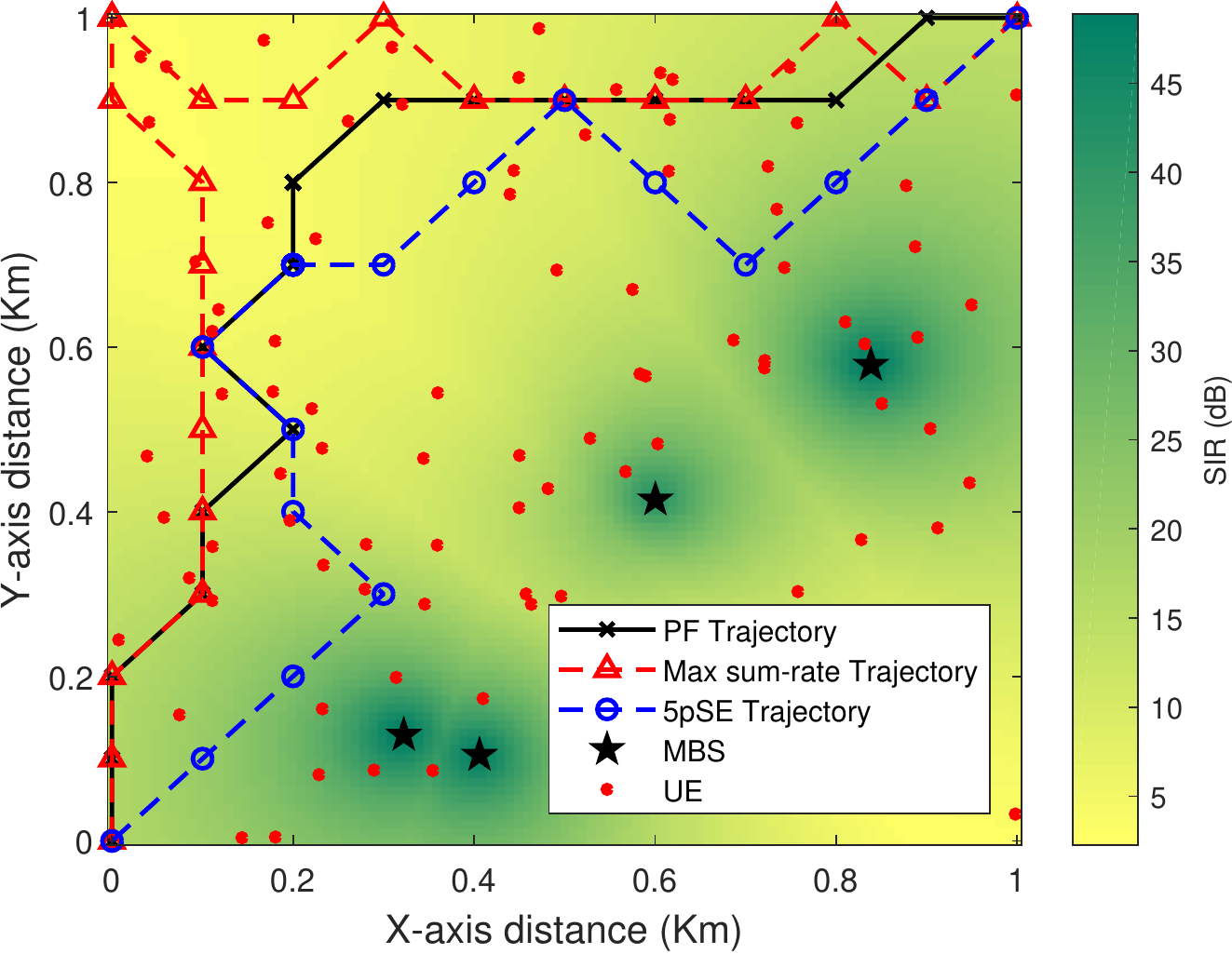}}%
			
		\subfloat[]{
			\includegraphics[width=0.8\linewidth]{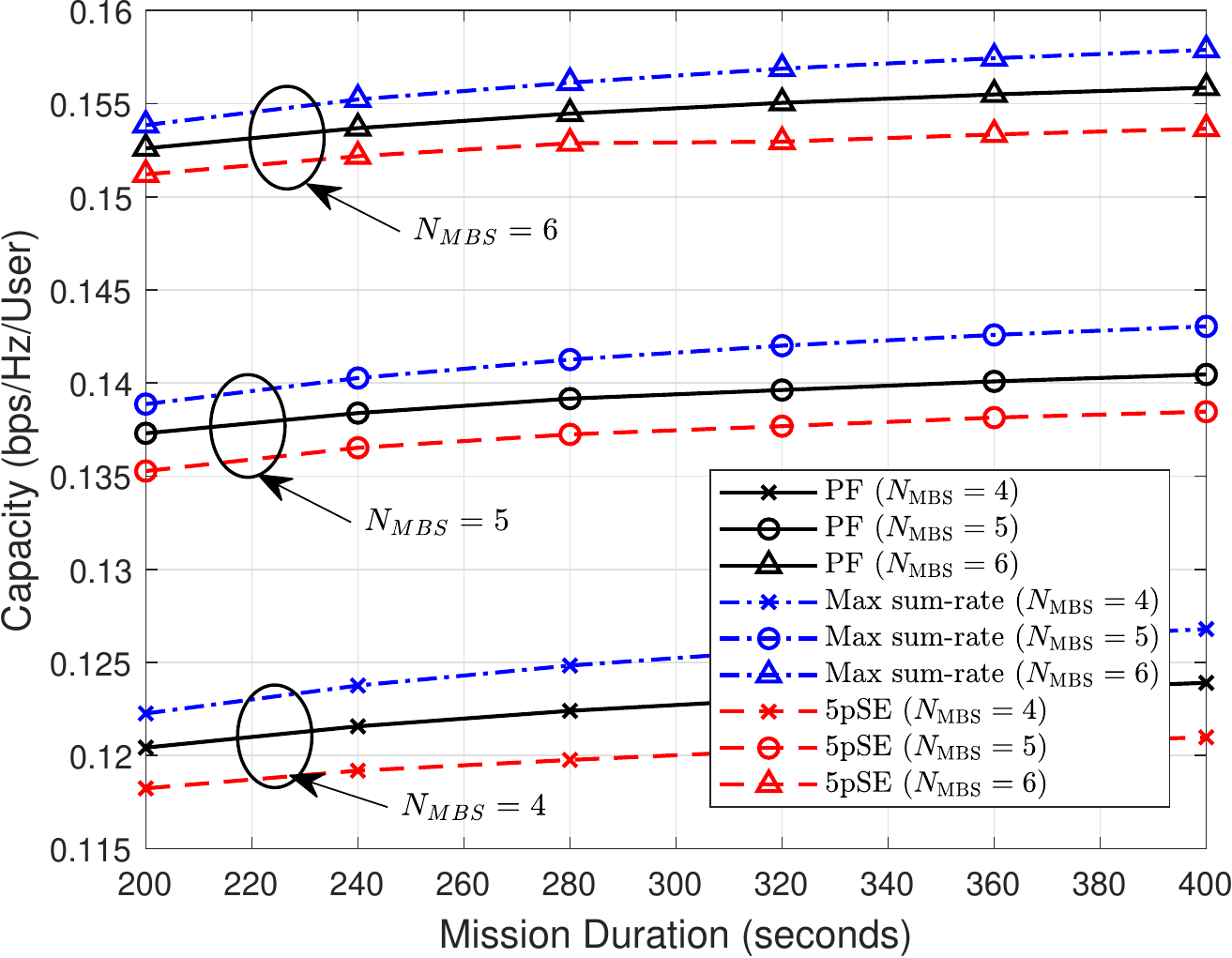}}%
\vspace{-.05cm}		
		\subfloat[]{
			\includegraphics[width=0.8\linewidth]{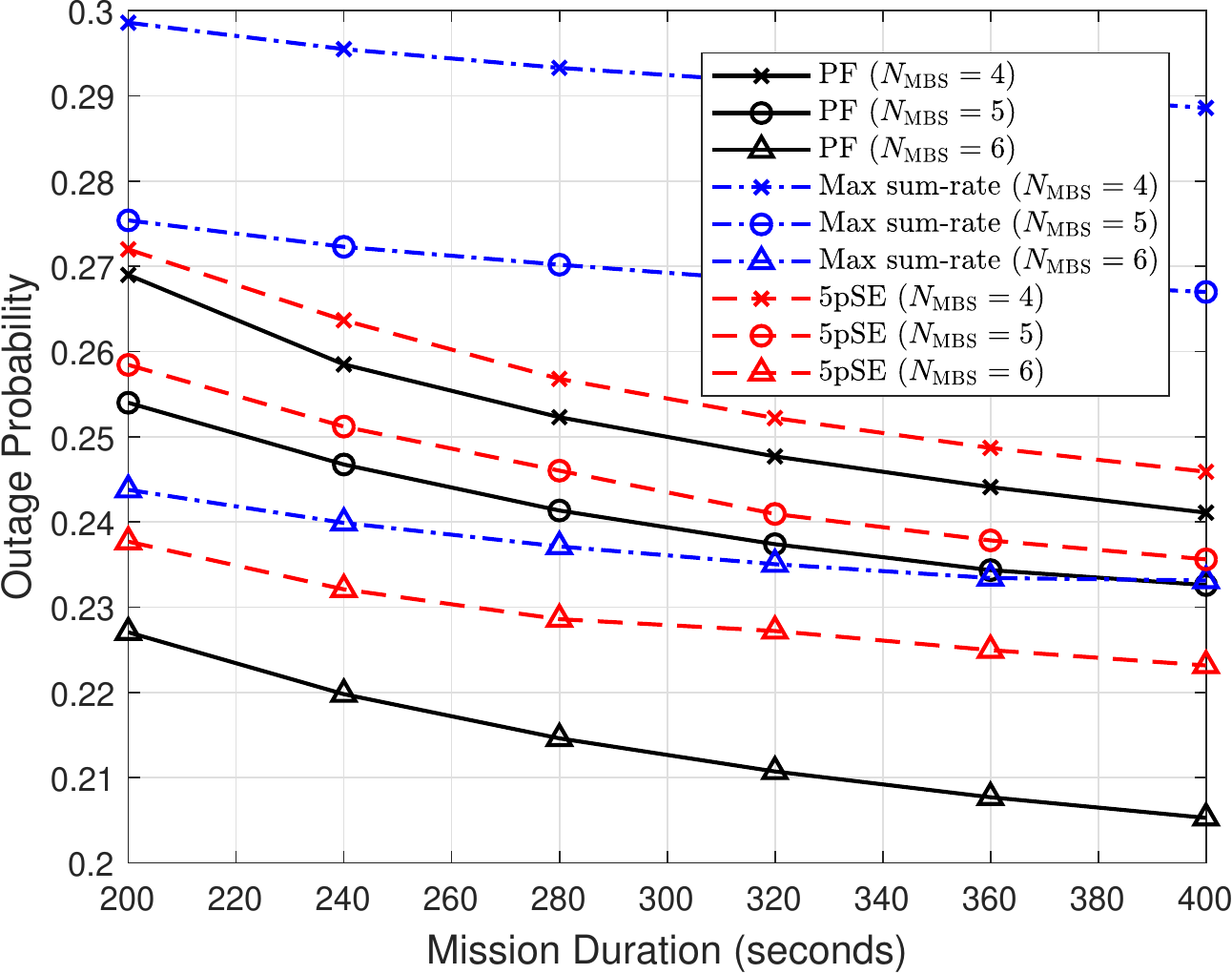}}%
		\caption {(a) Optimal trajectories of PF rate, Max sum-rate and 5pSE rate for $T=240$~s overlayed on SIR (dB) heat map at each discrete point.
		(b) Network per-user capacity comparison between PF rate, max sum-rate, and 5pSE trajectories. 
       (c) Outage Probability comparison between PF rate, max sum-rate, and 5pSE trajectories. 
        }
		\label{pf_and_max}
		\vspace{-3mm}
	\end{figure}

First, we investigate the trajectories associated with PF rate, Max sum-rate, and 5pSE rate, in a network with 4 MBSs and 100 UEs for $T=240$~s, starting from source $(0,0)$~km to destination $(1,1)$~km. Fig.~\ref{pf_and_max}(a) illustrates that, the optimal paths are clearly dependent on the SIR at each discrete point. The Max sum-rate trajectory tends move towards low SIR regions and tries to associate two or three UEs and provide downlink coverage to them. The PF rate trajectory tends to move in both high and low SIR regions to maintain a balance between rate and fairness. 5pSE trajectory on the other hand, associate some UEs with good throughput from the MBS, so that, the total resource of the MBS can be distributed between lesser number UEs which helps to improve the capacity of cell-edge UEs. Another interesting observation is that, while completing mission, the UAV tends to reach the optimal point (highest value among the 121 points) as soon as possible and stay there for a while before start moving towards the final destination to meet the time constraint for all three approaches. These points are (0.2, 0.8)~km, (0, 1)~km and (0.2, 0.7)~km for PF, Max sum-rate and 5pSE, respectively. This observation is consistent with \cite{rajeev}.

Next, we explore the capacity 
performance of different  trajectories by varying mission duration $T$ and number of MBSs ($N_{\mathrm{MBS}}$) in the network. We generated 20 networks and calculated optimal trajectories for each $T$ and for various $N_{\mathrm{MBS}}$. Fig.~\ref{pf_and_max}(b) illustrates the time-averaged per-UE capacity for different criteria. With the increasing mission duration, per UE capacity increases (due to possibility to reach out to further locations) and saturates as expected. The Max sum-rate trajectory outweighs the other two approaches in terms of per-user capacity. Higher $N_{\mathrm{MBS}}$ provides better SIR and hence, results in better per UE capacity which is also reflected here. 

Fig.~\ref{pf_and_max}(c) on the other hand depicts the outage probabilities associated with the different trajectories, where we consider that a UE is in outage if its rate is lower than $0.05$~bps/Hz. For this scenario, the PF trajectory provides better coverage probability than max-sum rate and 5pSE rate. Outage probability decreases with the increasing number of MBSs due to better coverage and SIR. The max sum-rate trajectory does not take deprived UEs into account and hence, provides the worst performance.

\begin{figure}[t]
		\centering
		\subfloat[]{
			\includegraphics[width=0.8\linewidth]{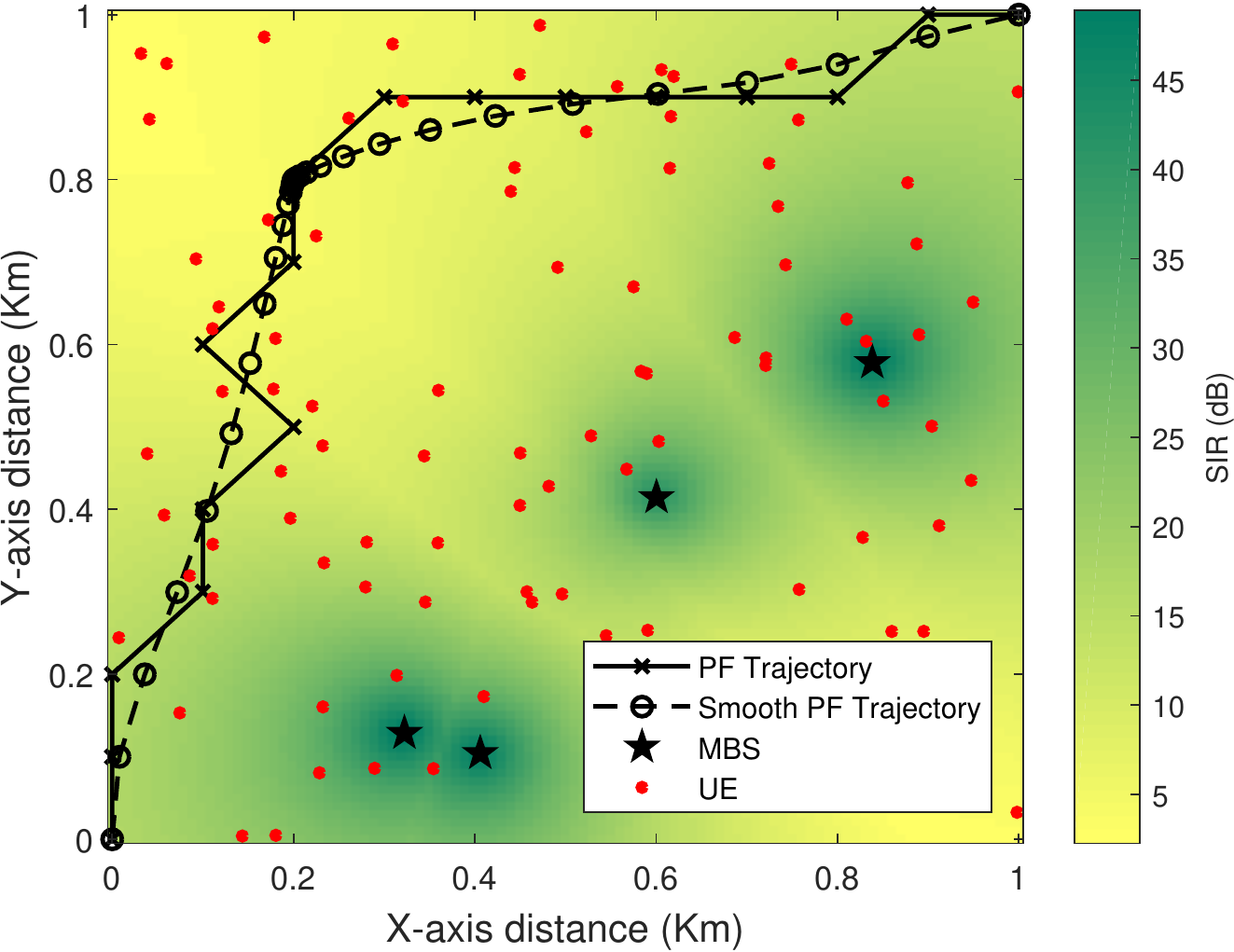}}%
			
		\subfloat[]{
			\includegraphics[width=0.8\linewidth]{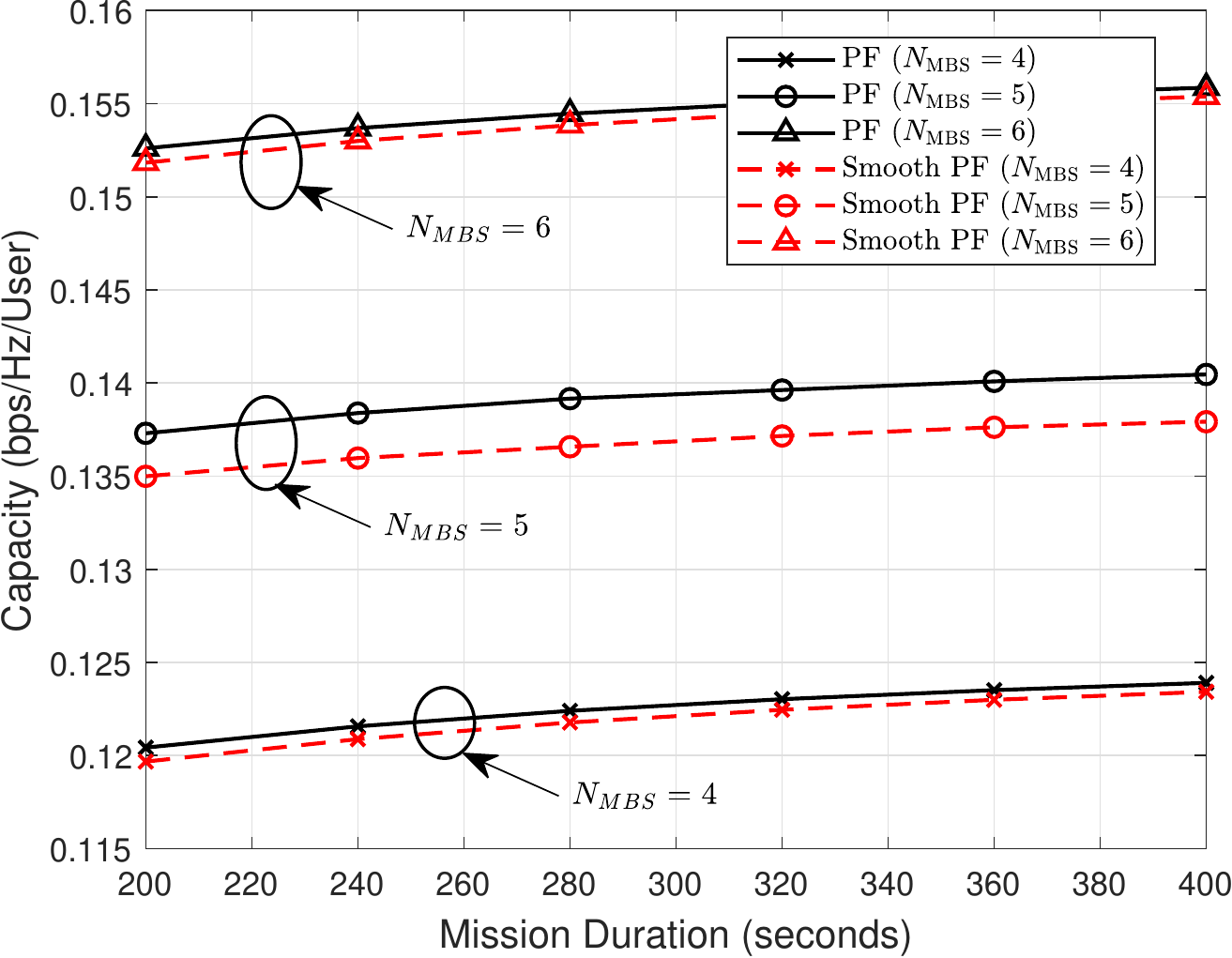}}%
\vspace{-.05cm}		
		\subfloat[]{
			\includegraphics[width=0.8\linewidth]{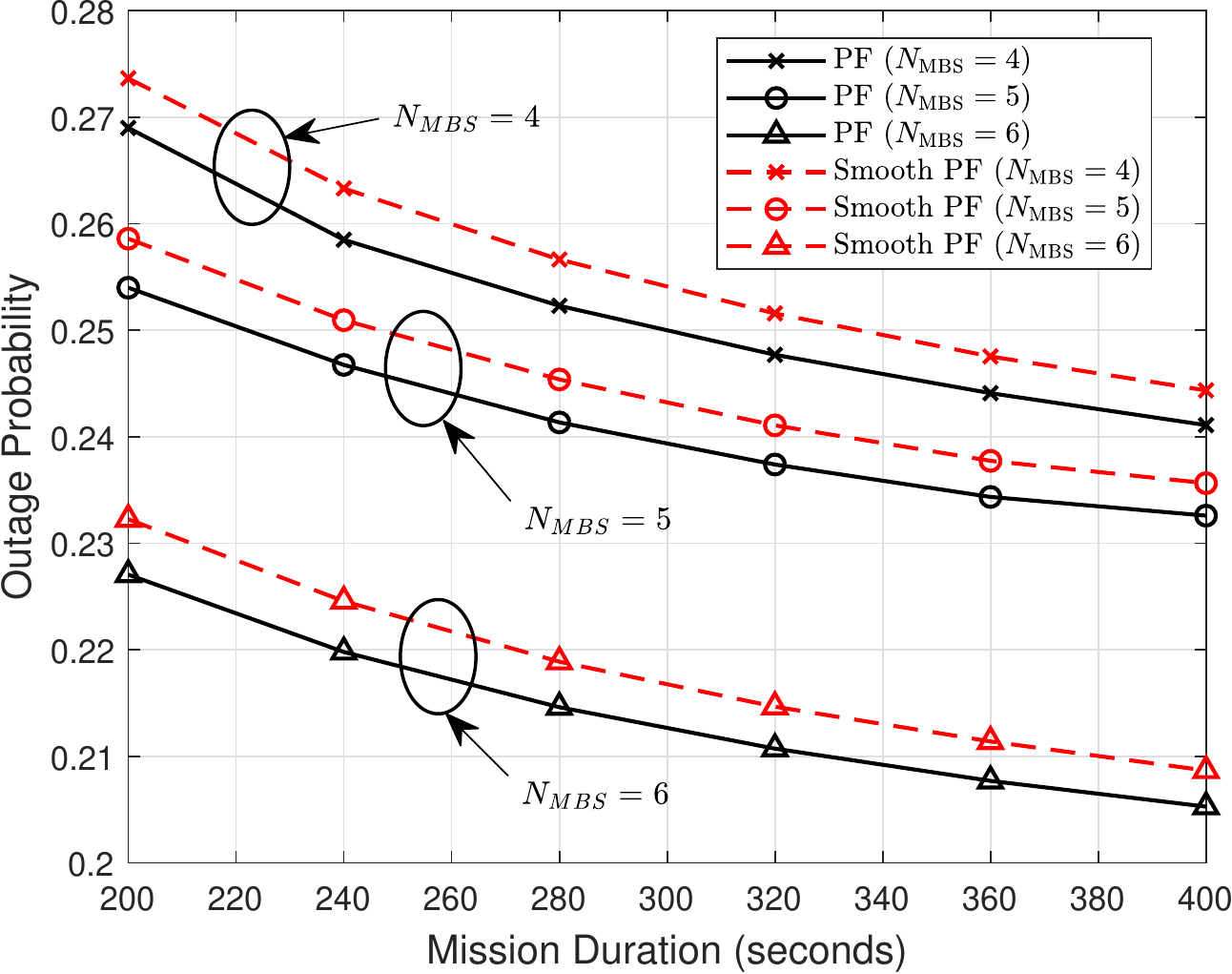}}%
		\caption {(a) Optimal trajectories of PF rate and smoothened PF-rate using Bezier curve for \textbf{$T=240$}~s overlapped on SIR (dB) heat map at each discrete point.
		(b) Network per-user capacity comparison between PF rate and smooth PF trajectory generated by the Bezier curve.
       (c) Outage probability comparison between PF rate, and smooth PF trajectory generated by the  Bezier curve. 
        }
		\label{pf_and_max2}
		\vspace{-3mm}
	\end{figure}
	
So far, the optimal paths are determined using DP algorithm. However, these paths contain only straight-line segments and sharp turns. Due to the kinematic and dynamic constraints of UAVs, these kind of paths cannot be followed in general. In fact, the UAV can maintain close to optimal performance while avoiding sharp turns by moving near the optimal points. We therefore explore Bezier curve to smooth the generated paths from DP technique. A Bezier curve is a parametric smoothing curve in 2D space which uses Bernstien polynomials to generate the basis. A Bezier curve of degree $n$ or order $n+1$ can be written as~\cite{bezier2}:
\begin{equation}
    b(\hat{t})=\sum_{i=0}^n \mathrm{\mathbf{P}_i}B_{i,n}(\hat{t}),\quad 0\leq \hat{t} \leq 1,
\end{equation}
where $\textbf{P}$ is the vector of $n+1$ control points and $B_{i,n}(\hat{t})$ are the Bernstein polynomials of degree $n$ which can explicitly expressed as, $ B_{i,n}(\hat{t})={n \choose i} (1-\hat{t})^{n-i}\hat{t}^i$.
  
We smooth the optimal trajectory for PF rate and plot in Fig.~\ref{pf_and_max2}(a). We can see that the smooth trajectory follows the original optimal path closely. It is worth noting that, the UAV tends to move very slowly near the optimal point to maintain the performance.

Next, we explore the capacity and outage performance comparison of PF rate trajectory and its pertinent smooth trajectory which is shown in Fig.~\ref{pf_and_max2}(b) and Fig.~\ref{pf_and_max2}(c), respectively. Our result show that the performance gap between these two trajectories is not significant. This is due to the fact that, the Bezier curve allows to smooth the trajectory to remain almost stalled near the optimal point. This helps the UAV to maintain good capacity and outage performance. Hence, we can use Bezier curve to smooth the discrete trajectories generated by using DP algorithm without any noticeable degradation in the overall network performance.

\section{Concluding Remarks}
In this paper, we presented the trajectory design problem in an interference prevalent downlink cellular network in order to maximize the PF rate, Max sum-rate, and 5pSE rate. We first formulate the trajectory optimization problems for different criteria, and solve them using the dynamic programming technique. We also explore and study the capacity and outage probability of the optimal trajectories. Our simulation results show that the PF rate trajectory provides better coverage performance while Max sum-rate provides the best per UE throughput. We then consider smoothing the sharp trajectories using Bezier curve and show the performance invariability of the smoothed trajectories. 


\bibliographystyle{IEEEtran} 
\bibliography{ref.bib}

\end{document}